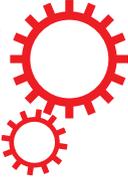

# Playing with universality classes of Barkhausen avalanches

Felipe Bohn[1], Gianfranco Durin[2,3], Marcio Assolin Correa[1], Núbia Ribeiro Machado[1], Rafael Domingues Della Pace[1], Carlos Chesman[1] & Rubem Luis Sommer[4]

Many systems crackle, from earthquakes and financial markets to Barkhausen effect in ferromagnetic materials. Despite the diversity in essence, the noise emitted in these dynamical systems consists of avalanche-like events with broad range of sizes and durations, characterized by power-law avalanche distributions and typical average avalanche shape that are fingerprints describing the universality class of the underlying avalanche dynamics. Here we focus on the crackling noise in ferromagnets and scrutinize the traditional statistics of Barkhausen avalanches in polycrystalline and amorphous ferromagnetic films having different thicknesses. We show how scaling exponents and average shape of the avalanches evolve with the structural character of the materials and film thickness. We find quantitative agreement between experiment and theoretical predictions of models for the magnetic domain wall dynamics, and then elucidate the universality classes of Barkhausen avalanches in ferromagnetic films. Thereby, we observe for the first time the dimensional crossover in the domain wall dynamics and the outcomes of the interplay between system dimensionality and range of interactions governing the domain wall dynamics on Barkhausen avalanches.

Crackling noise arises in many systems; when driven slowly, they respond with the emission of a noise consisting of series of sudden avalanche-like events with broad range of sizes and durations[1–4]. In the past decade or so, studies on avalanche dynamics and crackling noise have uncovered an underlying criticality in a wide variety of fundamentally different systems that show strikingly similar behavior, e.g., earthquakes[1], plastic deformations[5,6], microfractures[7], sheared granular materials[8], vortices in superconductors[9], dimming events in stars[10], and financial markets[11]. However, until now, the most striking, paradigmatic example of self-organization and non-equilibrium critical dynamics is undoubtedly the complex microscopic magnetization process through the jerky motion of magnetic domain walls (DW) in ferromagnetic materials[3]. In the presence of a smooth, slow-varying external magnetic field, the material responds through a sequence of discrete and irregular jumps of magnetization (as we can see in Fig. 1(a)), known as Barkhausen effect[12–15], which can be detected as a crackling noise (Fig. 1(b)) in a suitable experimental setup.

The critical dynamics across all these systems is characterized by avalanches with scale-invariant properties, power-law distributions, and universal features — it means that a same behavior will be shared among large family of materials and even different systems that are in the same universality class, whereas the behavior will typically differ between systems that are fundamentally different[1–4]. The power-law scaling exponents and the typical avalanche shape emerge as fingerprints describing the universality class of the underlying avalanche dynamics[1]. Hence, the Barkhausen avalanches recorded in ferromagnetic materials are in this context a wonderful playground for investigating scaling phenomena found in the most diverse systems exhibiting crackling noise, providing hints on this exciting, still-evolving field.

The universality class of Barkhausen avalanches in a sample is usually identified by measuring the distributions of avalanche sizes and durations, the joint distribution of sizes and durations, and the average temporal avalanche shape. Much efforts have been devoted to link noise statistics to characteristic features of the materials. Despite Barkhausen avalanches have been investigated experimentally for decades in bulk materials[16–31] and films[32–49], universality was questioned for a long time. An important step towards understanding Barkhausen avalanches has been achieved by Durin and Zapperi[31], who first provided consistent interpretation of the Barkhausen statistics in bulk materials, well-known systems exhibiting three-dimensional magnetic behavior. From classical inductive experiments, the scaling exponents associated with such distributions have been found different according to the structural character of the sample, placing polycrystalline and amorphous materials in two distinct universality classes differing in the kind and range of interactions governing the DW dynamics. For films in turn, universality is still under debate. The Barkhausen avalanches have been investigated primarily through magneto-optical techniques[32–42] and, just more recently, with the inductive technique[43–49]. These first reliable magneto-optical experiments have shown that the magnetic behavior in thin films typically differs from that found for bulk materials, a characteristic entirely devoted to the dimensionality of the system. A step forward in the subject has been given by Ryu and colleagues[37], who addressed the scaling behavior of Barkhausen criticality in a ferromagnetic 50-nm-thick MnAs thin film, a system with essentially two-dimensional magnetization dynamics due to the reduced thickness. From sophisticated magneto-optical observations of the avalanches, the scaling behavior has been experimentally tuned by varying the temperature close but below the Curie temperature of this given film. The modification of a single scaling exponent, taking place simultaneously to a change in the DW morphology, discloses a crossover between two distinct universality classes, which is caused by the competition between long-range dipolar interaction and the short-range DW surface tension. However, it worth noting that this is not the whole story. In the last years, our group[43–47] has explored the Barkhausen noise from inductive experiments, bringing to light the scaling exponents of the avalanche distributions and the average avalanche shape for polycrystalline and amorphous thicker films. Strikingly, the results of this wide statistical treatment of Barkhausen avalanches, besides corroborating the universality classes found for bulk materials, suggest that the two-dimensional DW dynamics is not shared among films in all thickness ranges.

[1]Departamento de Física, Universidade Federal do Rio Grande do Norte, 59078-900, Natal, RN, Brazil. [2]Istituto Nazionale di Ricerca Metrologica, Strada delle Cacce 91, 10135, Torino, Italy. [3]ISI Foundation, Via Alassio 11/c, 10126, Torino, Italy. [4]Centro Brasileiro de Pesquisas Físicas, Rua Dr. Xavier Sigaud 150, Urca, 22290-180, Rio de Janeiro, RJ, Brazil. Correspondence and requests for materials should be addressed to F.B. (email: felipebohn@fisica.ufrn.br)

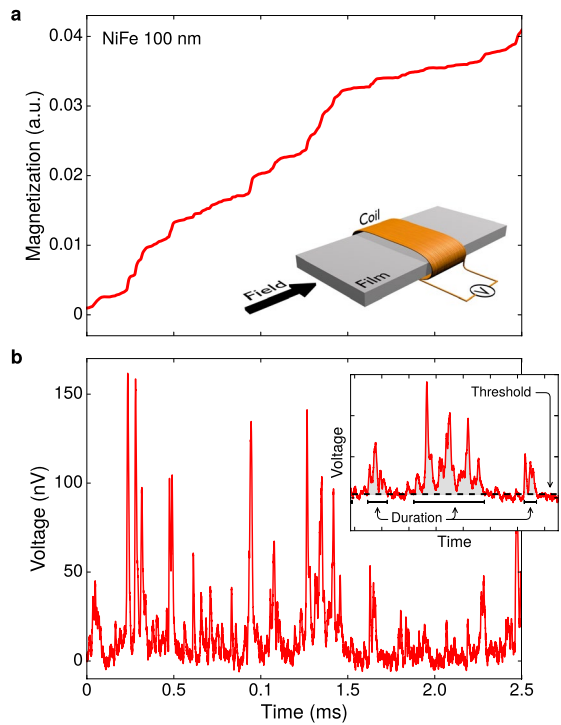

**Figure 1.** Magnetization jumps and the Barkhausen effect. (**a**) Magnetization curve, as a function of the time, of a 100-nm-thick ferromagnetic NiFe film submitted to a smooth, slow-varying external magnetic field. The magnification of the curve reveals that the change in magnetization is not smooth, but exhibits discrete and irregular jumps. The jumps of magnetization are due to the jerky motion of the magnetic domain walls in a disordered medium, a result of the interactions between DW and pinning centers, such as defects, impurities, dislocations, and grain boundaries. In a typical Barkhausen noise experiment, the changes of magnetization are detected by a pickup coil wound around the ferromagnetic material. As the magnetization changes, the respective variation of the magnetic flux induces a voltage signal in the coil that can be amplified and recorded. (**b**) The crackling response in magnetic systems is the Barkhausen noise, which itself consists in the time series of voltage pulses detected by the pickup coil. Notice that the Barkhausen noise shown in (b) is proportional to the time derivative of the magnetization in (a). The noise in correspondence to the magnetization jumps is a series of Barkhausen avalanches with broad range of sizes and durations. The inset shows an example of how the avalanches are extracted. A threshold (dashed line) is set to properly define the beginning and end of each Barkhausen avalanche. Three different avalanches are denoted here by the gray zones. The duration of the example avalanches is marked by solid intervals. The duration $T$ is thus estimated as the time interval between the two successive intersections of the signal with the threshold. The area underneath the avalanche signal, between the same points, is defined as the avalanche size $s$.

Nowadays, generally, the scaling behavior of Barkhausen avalanches is understood in terms of depinning transition of domain walls[30]. Remarkably, experimental investigations confronted to theoretical predictions and simulations have uncovered that scaling exponents and average shape of Barkhausen avalanches reflect fundamental features of the underlying magnetization dynamics, as system dimensionality and kind and range of interactions governing the DW motion[12,13]. But despite the recent advances in the field, in contrast to bulk materials, our understanding on Barkhausen avalanches in films is far from complete. Specifically, due to experimental difficulties and scarce statistical data, the influence of structural character and film thickness on the scaling behavior is an open question. So a general framework for the universality classes of Barkhausen avalanches in films is still lacking.

|  | Two-dimensional systems | | | Three-dimensional systems | | |
|---|---|---|---|---|---|---|
|  | $\tau$ | $\alpha$ | $1/\sigma\nu z$ | $\tau$ | $\alpha$ | $1/\sigma\nu z$ |
| Long-range interactions | $\simeq$1.33[52–54] | $\simeq$1.5[52–54] | $\simeq$1.5[52–54] | 1.50[29–31] | 2.0[29–31] | 2.0[29–31] |
| Short-range interactions | $\simeq$1.06[51,54] | — | — | 1.27[29–31,50,51] | 1.5[29–31] | 1.77[29–31] |

**Table 1.** The scaling exponents predicted by theoretical models and simulations for two- and three-dimensional systems, with long- and short-range interactions governing the DW dynamics. Notice that some exponents are not shown, exposing that even in the theoretical side many questions in the field remain to be solved.

Here we report an experimental study of the statistics of Barkhausen avalanches in ferromagnetic films and show how scaling exponents and average shape of the avalanches evolve with the structural character of the materials and film thickness. By comparing our experiments with theoretical predictions of models for the DW dynamics, we interpret the universality classes of Barkhausen avalanches in ferromagnetic films. Thereby, we provide an experimental evidence for the dimensional crossover in the DW dynamics, and disclose outcomes of the interplay between system dimensionality and range of interactions governing the DW dynamics on Barkhausen avalanches.

## Results

**Scrutinizing Barkhausen avalanches in films.** We systematically analyze the statistics of Barkhausen avalanches in polycrystalline and amorphous ferromagnetic films with thicknesses from 20 to 1000 nm (see Methods for details on the films, experiments and statistical analysis of the avalanches). Having established a sophisticated method of extraction of the avalanches due the low intensity of the signal (see Methods and Fig. 1(b)), we obtain a wide statistical analysis measuring the distributions of avalanche sizes and durations, the joint distribution of sizes and durations, the power spectrum, and the average avalanche shape. So, we probe for the influence of the structural character of the materials and film thickness on the DW dynamics, and play with universality classes of Barkhausen avalanches in an experimentally controlled manner.

**Scaling exponents and the average avalanche shape.** Theoretical models has always been crucial in the broad field of crackling noise. Quantitative comparison between experiment and predictions is primary done through scaling exponents. This means that if the theory correctly describes an experiment, the exponents will agree[1]. Here we consider three exponents $\tau$, $\alpha$, and $1/\sigma\nu z$ (See Methods). In the scaling regime, these are defined from $P(s) \sim s^{-\tau}$, $P(T) \sim s^{-\alpha}$, and $\langle s \rangle \sim T^{1/\sigma\nu z}$. Specifically for ferromagnetic films, the key to the understanding of the Barkhausen avalanche statistics resides in the *interplay* between system dimensionality and range of interactions. Many approaches capturing essential features of the magnetic systems have been developed to mimic the DW dynamics. So, to interpret our experimental results, we summarize in Table 1 the scaling exponents predicted for two- and three-dimensional systems, with long- and short-range interactions governing the DW dynamics[29–31,50–54].

Going beyond power laws, the average avalanche shape characterized by universal scaling functions is a sharper tool to identify universality classes[1]. There are two types of averages that can be performed to find different universal profiles. The first is the average temporal avalanche shape $\langle V(t|T) \rangle$, obtained averaging over avalanches of a given duration, whereas the second type is the average avalanche shape for a specific size, $\langle V(S|s) \rangle$, involving an average over avalanches of the same size; both avalanche shapes follow scaling forms dependent on the universality class through the scaling exponent $1/\sigma\nu z$. So, we also look at the avalanche shape and compare our experimental results with the recent theoretical advances achieved by Laurson *et al.*[4] (See Methods).

**Polycrystalline films.** Figure 2 shows the Barkhausen avalanche statistics for the polycrystalline films having different thicknesses, and Table 2 presents the measured scaling exponents. For all thicknesses, the distributions in Fig. 2(a–c) show cutoff-limited power-law scaling behavior, revealing genuine scale invariance. The power laws with cutoffs are understood as a fingerprint of a critical behavior of the magnetization process[1]. The most noticeable feature related to the power-law behavior is that the scaling exponents vary as the film thickness is reduced from 100 to 50 nm. Different sets of exponents support the idea that there are distinct kinds of behaviors, the universality classes. Here we clearly see that the polycrystalline films split into two universality classes. The first class includes films with thicknesses above 100 nm, characterized by exponents $\tau \approx 1.50$, $\alpha \approx 2.0$, and $1/\sigma\nu z \approx 2.0$ measured for the smallest magnetic field frequency. These results are also shown and discussed in detail in ref.[45]. For the films in this first class, we observe well-known rate effects, including the frequency dependence of $\tau$ and $\alpha$, in agreement with earlier findings for bulk polycrystalline materials[31]. Moreover, through the comparison between experimental and theoretical exponents, we find that these films exhibit critical behavior consistent with the mean-field theory describing three-dimensional magnets, which predicts $\tau = 1.50$, $\alpha = 2.0$, and $1/\sigma\nu z = 2.0$. It discloses in polycrystalline films thicker than 100 nm a typical three-dimensional DW dynamics governed by long-range dipolar interactions[29–31], as we can see from Table 1. In contrast to these thickest films, the films thinner than 50 nm belong to the second universality class, characterized by frequency-insensitive exponents $\tau \approx 1.33$, $\alpha \approx 1.5$, and $1/\sigma\nu z \approx 1.6$. It is worth remarking that a similar experimental $\tau$ value has previously been reported for different crystalline films with thicknesses below 50 nm[35–39]. Moreover, on the theoretical side, we verify that the exponents are in quite-well concordance with the set of values $\tau \simeq 1.33$, $\alpha \simeq 1.5$, and $1/\sigma\nu z \simeq 1.5$ (See Table 1). So, the agreement of experimental results with theoretical predictions and simulations reveals that polycrystalline films thinner than 50 nm have an universal two-dimensional DW dynamics dominated by long-range dipolar interactions[52–54].

| | Film | $\tau$ | $\alpha$ | $1/\sigma\nu z$ |
|---|---|---|---|---|
| Polycrystals | NiFe 20 nm | 1.33 ± 0.05 | 1.50 ± 0.07 | 1.54 ± 0.05 |
| | NiFe 50 nm | 1.32 ± 0.05 | 1.56 ± 0.08 | 1.63 ± 0.07 |
| | NiFe 100 nm | 1.51 ± 0.03 | 1.99 ± 0.06 | 1.95 ± 0.04 |
| | NiFe 200 nm | 1.49 ± 0.02 | 1.94 ± 0.04 | 1.91 ± 0.03 |
| | NiFe 500 nm | 1.45 ± 0.02 | 1.94 ± 0.05 | 2.09 ± 0.05 |
| | NiFe 1000 nm | 1.43 ± 0.04 | 1.90 ± 0.07 | 2.08 ± 0.05 |
| Amorphous | B4 50 nm | 1.33 ± 0.06 | 1.50 ± 0.07 | 1.50 ± 0.04 |
| | FeSiB 50 nm | 1.35 ± 0.06 | 1.55 ± 0.07 | 1.57 ± 0.03 |
| | B4 100 nm | 1.28 ± 0.02 | 1.52 ± 0.05 | 1.82 ± 0.05 |
| | FeSiB 100 nm | 1.30 ± 0.05 | 1.53 ± 0.08 | 1.80 ± 0.07 |
| | CoFe 100 nm | 1.30 ± 0.05 | 1.53 ± 0.08 | 1.79 ± 0.06 |
| | B9 150 nm | 1.28 ± 0.04 | 1.52 ± 0.05 | 1.82 ± 0.05 |
| | B6 200 nm | 1.28 ± 0.03 | 1.50 ± 0.07 | 1.76 ± 0.08 |
| | FeSiB 500 nm | 1.26 ± 0.03 | 1.48 ± 0.04 | 1.80 ± 0.03 |
| | CoSiB 1000 nm | 1.27 ± 0.02 | 1.50 ± 0.04 | 1.81 ± 0.05 |

**Table 2.** The experimental scaling exponents for polycrystalline and amorphous films having different thicknesses, measured with the smallest driving magnetic field frequency, 50 mHz.

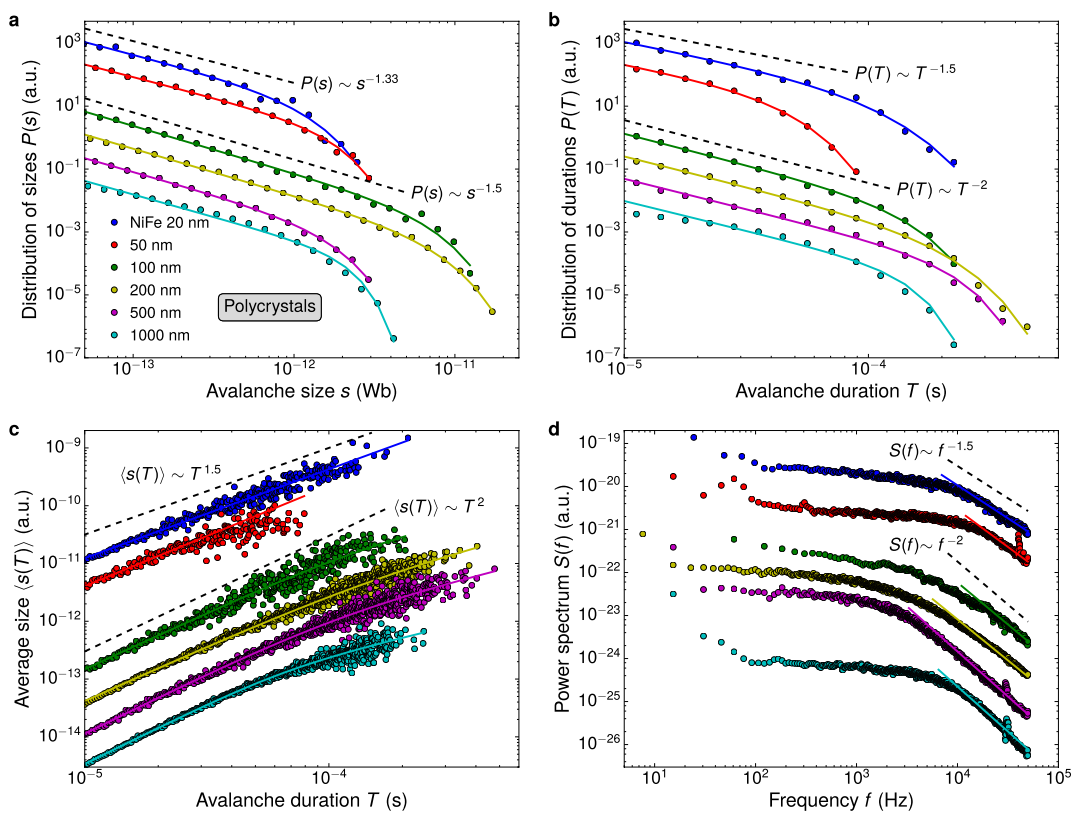

**Figure 2.** Dimensional crossover in the DW dynamics. Statistical analysis of Barkhausen avalanches in polycrystalline NiFe films having different thicknesses, from 20 to 1000 nm. (**a**) Distributions of avalanche sizes measured at the smallest driving field frequency, 50 mHz. The solid lines are cutoff-limited power-law fittings obtained with Eq. (1). (**b**) Similar plot for the distributions of avalanche durations, with fittings obtained using Eq. (2). (**c**) Average size as a function of the avalanche duration, with fittings obtained using Eq. (3). (**d**) The power spectra. Here, the solid lines are power laws obtained using Eq. (4) with slopes $1/\sigma\nu z$, the exponent measured from the relationship between $\langle s \rangle$ and $T$ for each film. In (a–d), the data are vertically shifted for clarity. The dashed lines are power laws whose slopes correspond to the exponents of the two universality classes found for polycrystalline films. In particular, the experimental results for the universality class that includes the thickest films are also found and discussed in ref.[45].

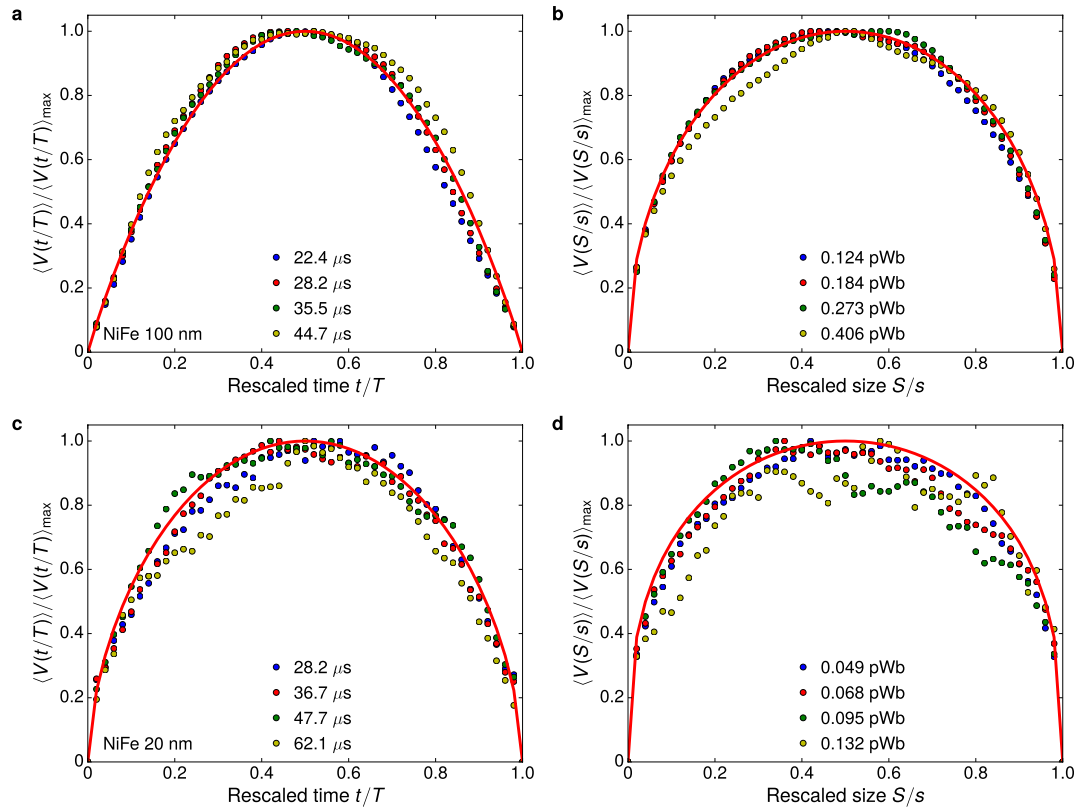

**Figure 3.** Evolution of the avalanche shape with the universality class. (**a**) Temporal average avalanche shape for different avalanche durations $T$, rescaled to unit height and duration, and (**b**) average avalanche shape for fixed avalanche sizes $s$, rescaled to unit height and size, both for the polycrystalline NiFe film with thickness of 100 nm, a three-dimensional system with long-range interactions. The symbols are experimental data for different durations or sizes of the avalanches, whereas the solid line is the correspondent theoretical prediction, given by Eqs (5) or (6), obtained with $1/\sigma\nu z = 1.95$ measured from the relationship between $\langle s \rangle$ and $T$. (**c**,**d**) Similar plots for the polycrystalline 20-nm-thick NiFe film, a two-dimensional system with long-range interactions, having $1/\sigma\nu z = 1.54$.

An important test of consistency with theoretical predictions is provided by the exponent relation $(\alpha - 1)/(\tau - 1) = 1/\sigma\nu z$[55]. We verify that the exponents within the measurement error satisfy this equation for all thicknesses. Yet we observe that the power spectrum in Fig. 2(d) follows a power-law behavior at the range of high frequencies. Thus, we also confirm another theoretical prediction, $S(f) \sim f^{-1/\sigma\nu z}$, corroborating that the very same exponent may describe the scaling regime in the power spectrum and the power-law relationship between $\langle s \rangle$ and $T$. Moreover, it is interesting to notice the remarkable stability of the scaling exponents within each universality class. Specifically, the exponents have similar values despite the magnetic properties, including magnetic domain structure, magnetic anisotropy and permeability, as well as density of defects, stress level, and the own thickness are changing simultaneously[45,56]. This result is consistent with theoretical studies predicting that micro and macroscopic details of the material do not affect the exponents, but only alter the cutoff[1,12,13]. In particular, a straight consequence of the interplay of all these changes is that no systematic variation of the cutoff with thickness is found.

Further, we focus on the measurement of the average avalanche shape. Figure 3 presents the avalanche shapes for both thick and thin polycrystalline films as representative results for the two universality classes. Notice the striking agreement between experiment and theoretical predictions, including three important features: symmetry of the shapes, the exponent $1/\sigma\nu z$, and the scaling function. By employing films, retardation effects due to eddy currents are suppressed by the sample geometry[44]. It avoids the familiar leftward asymmetry found for bulk materials, yielding symmetric avalanche shapes[44]. It is worth noting that two well-known predictions for mean-field systems are retrieved here: $\langle V(t|T) \rangle$ is described in the scaling regime by an inverted parabola, and $\langle V(S|s) \rangle$ is given by a semicircle[13,27]. Both are found for the films thicker than 100 nm, whose $1/\sigma\nu z \approx 2$ in the scaling regime, as we can see in Fig. 3(a,b). For thinner films in Fig. 3(c,d) though, whose $1/\sigma\nu z \approx 1.6$, the average avalanche shapes differ from the mean-field forms. These findings disclose that the average shapes of the avalanches evolve with the universality class, and are perfectly described by the general scaling forms reported in ref.[4], both in and beyond mean field.

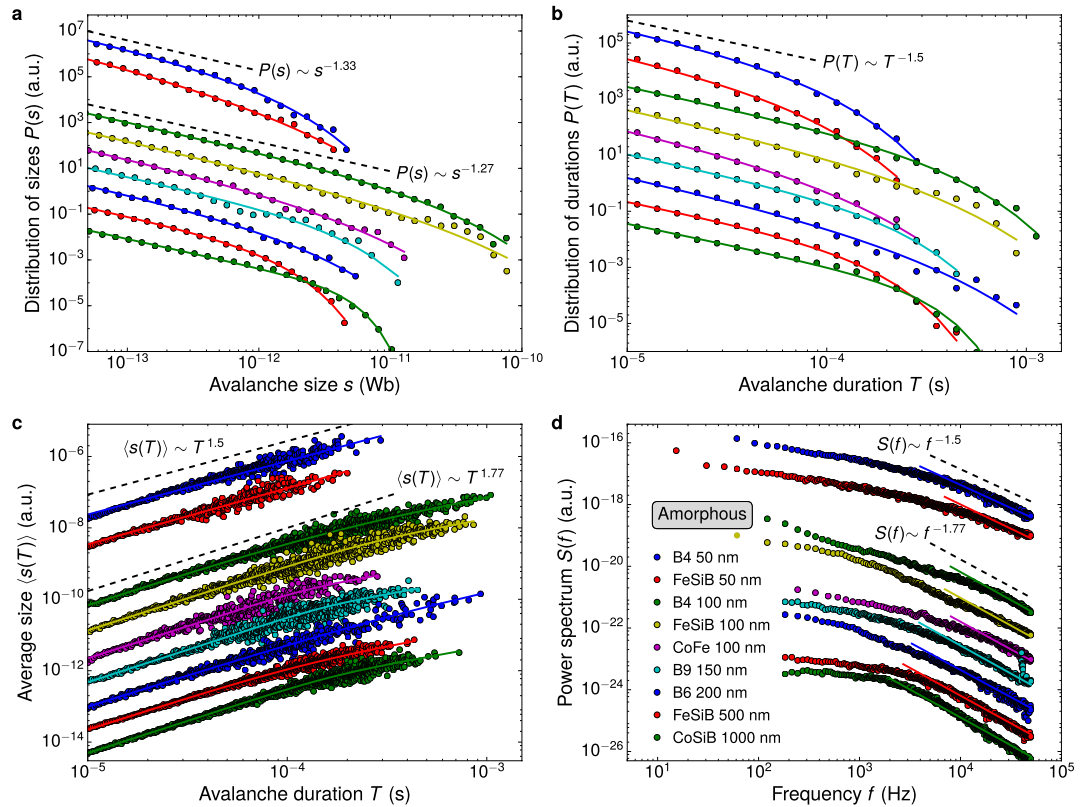

**Figure 4.** Interplay between system dimensionality and range of interactions governing the DW dynamics. Statistical analysis of Barkhausen avalanches in films of amorphous alloys with different thicknesses. (**a**) Distributions of avalanche sizes, (**b**) distributions of avalanche durations, (**c**) average size as a function of the avalanche duration, and (**d**) power spectra measured at the smallest driving field frequency, 50 mHz. For the distributions, the solid lines are the correspondent cutoff-limited power-law fittings obtained using Eqs (1), (2) and (3), while for the power spectrum, the solid lines are power laws obtained using Eq. (4) with slopes $1/\sigma\nu z$ for each film. In (a–d), the data are vertically shifted for clarity. The dashed lines are power laws whose slopes correspond to the exponents of the two universality classes found for amorphous films. In particular, the experimental results for the FeSiB films with thickness of 100 and 500 nm are also found in ref.[46]. Nevertheless, notice here the robustness of the scaling behavior for each universality class. This behavior is clearly not affected by the composition of the films.

**Amorphous films.** Figure 4 shows the dependence with thickness of the Barkhausen avalanche statistics for amorphous films, while Table 2 presents the measured scaling exponents. Similarly to polycrystals, the cutoff-limited power-law scaling behavior in the distributions of Fig. 4(a–c) and the power-law in the power spectrum of Fig. 4(d) are found for amorphous films. It is noteworthy that, as a test of consistency with theoretical predictions, we confirm that the $\tau$ and $\alpha$, and $1/\sigma\nu z$ measured for all thicknesses also satisfy the equation relating these three exponents[55].

Curiously enough, at first glance, the exponents $\tau$ and $\alpha$ might mislead us, suggesting a common critical behavior for all amorphous films, irrespective of the thickness and composition. Moreover, notice in Table 1 that two universality classes have very similar exponents $\tau$ and $\alpha$. So theoretical predictions for both exponents would lead us to two controversial interpretations, raising doubts on the underlying critical behavior. However, despite $\tau$ and $\alpha$ behave in a remarkably similar manner, a closer examination of the exponents, including $1/\sigma\nu z$, shows us that the amorphous films split into two distinct universality classes too. Indeed, the scaling relation between $\langle s \rangle$ and $T$ is known as a robust quantity, and a reliable test to identify universality classes[25]. For all amorphous films, no field frequency dependence of the exponents is found. The first universality class includes films with thicknesses above 100 nm and is characterized by exponents $\tau \approx 1.28$, $\alpha \approx 1.5$, and $1/\sigma\nu z \approx 1.8$, values comparable with those previously reported for bulk amorphous materials[31]. By the way, these results for the FeSiB films in this class are also found in ref.[46], but we recall them here to reinforce the robustness of the scaling behavior, corroborating that it is not affected by the composition of the films as well as by the strong modifications of the magnetic properties and magnetic domain structure taking place within this thickness range[43,46,57]. In addition, the values of the exponents in the first universality class are compatible with $\tau = 1.27$, $\alpha = 1.5$, and $1/\sigma\nu z = 1.77$ (See Table 1), predictions of models in which dipolar interactions are neglected in the DW motion[29–31,50,51], as expected for amorphous materials[31]. Therefore, the exponents suggest that amorphous films thicker 100 nm present a three-dimensional magnetic behavior with short-range DW surface tension governing the DW dynamics. Next, amorphous films thinner than 50 nm, also irrespective of the composition, are found in the second universality class, characterized by the exponents $\tau \approx 1.33$, $\alpha \approx 1.5$, and $1/\sigma\nu z \approx 1.55$. Here, it is very interesting to note that amorphous and polycrystalline thin films have similar exponents within the measurement error (See Table 2), suggesting they share the very same DW dynamics. The exponents are in quantitative agreement with $\tau \simeq 1.33$, $\alpha \simeq 1.5$, and $1/\sigma\nu z \simeq 1.5$[52–54] shown in Table 1. As a straight consequence, we find that amorphous films thinner than 50 nm also present a two-dimensional DW dynamics dominated by long-range interactions of dipolar origin.

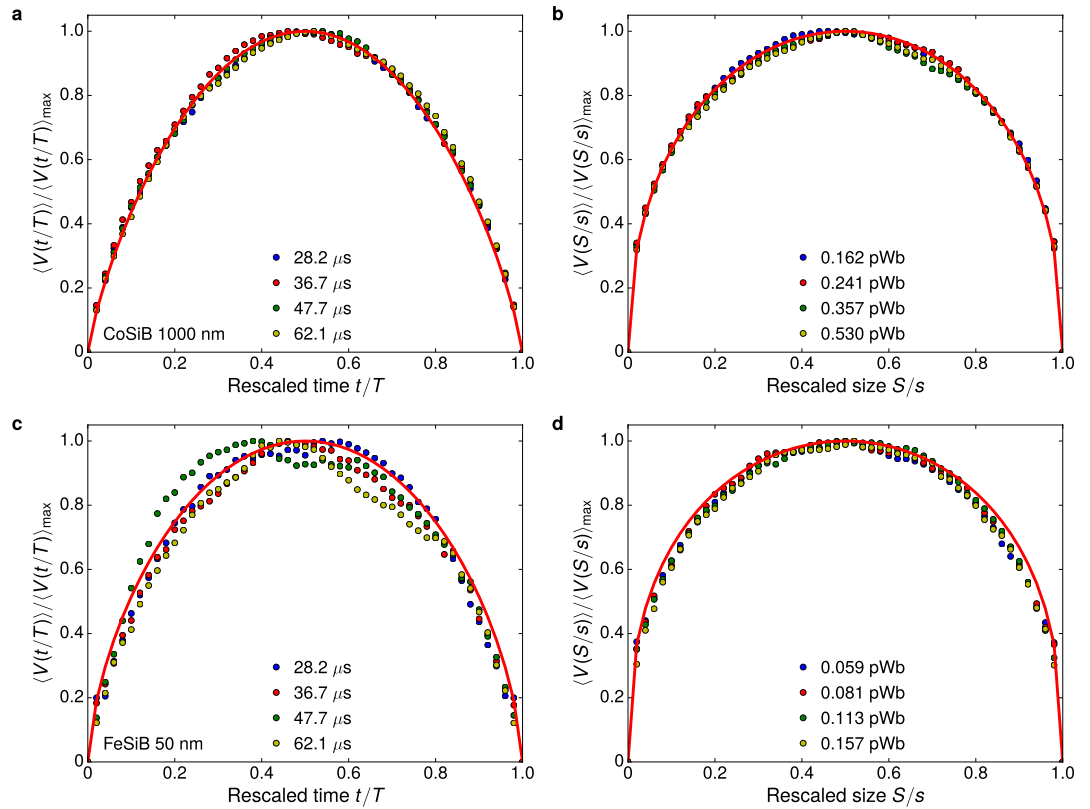

**Figure 5.** Universality classes beyond power laws and the still-evolving avalanche shape. (**a**) Temporal average avalanche shape for different avalanche durations $T$, rescaled to unit height and duration, and (**b**) average avalanche shape for fixed avalanche sizes $s$, rescaled to unit height and size, both for the amorphous CoSiB film with thickness of 1000 nm, a three-dimensional system with short-range interactions. The symbols are experimental data for different durations or sizes of the avalanches, whereas the solid line is the correspondent theoretical prediction, given by Eqs (5) or (6), obtained with $1/\sigma\nu z = 1.81$ measured from the relationship between $\langle s \rangle$ and $T$. (**c,d**) Similar plots for the amorphous FeSiB film with thickness of 50 nm, a two-dimensional system with long-range interactions, having $1/\sigma\nu z = 1.57$.

Last but not least, Fig. 5 presents the avalanche shapes for selected thick and thin amorphous films as representative results for the two universality classes. We clearly see that the average avalanche shapes evolve with the universality class, as expected[4]. Noticeably, experiment and theory again agree quite well, including features as the symmetry of the shapes due to absence of eddy current effects[44] and the scaling form ascribed to $1/\sigma\nu z$[4]. Thus, we corroborate the exponent estimated from the joint distribution of sizes and durations, as well as we also confirm the collapse and form of average avalanche shapes as a powerful alternative way to estimate the exponent $1/\sigma\nu z$.

## Discussion

Our findings raise interesting issues on the universality classes of Barkhausen avalanches. By comparing our experiments with theoretical predictions of models for the DW dynamics, we find that polycrystalline and amorphous films with distinct thicknesses assume values consistent with three well-defined universality classes. Specifically, the films split into the following classes of materials: (*i*) Polycrystalline films thicker than 100 nm presenting three-dimensional DW dynamics governed by long-range dipolar interactions; (*ii*) Amorphous films thicker than 100 nm having three-dimensional magnetic behavior with short-range DW surface tension governing the DW dynamics; (*iii*) Polycrystalline and amorphous films thinner than 50 nm with a two-dimensional DW dynamics dominated by strong long-range dipolar interactions. As a consequence, the changes found in scaling exponents and avalanche shape indicate modifications in the critical behavior of the system, i.e., the system passes from one universality class to another.

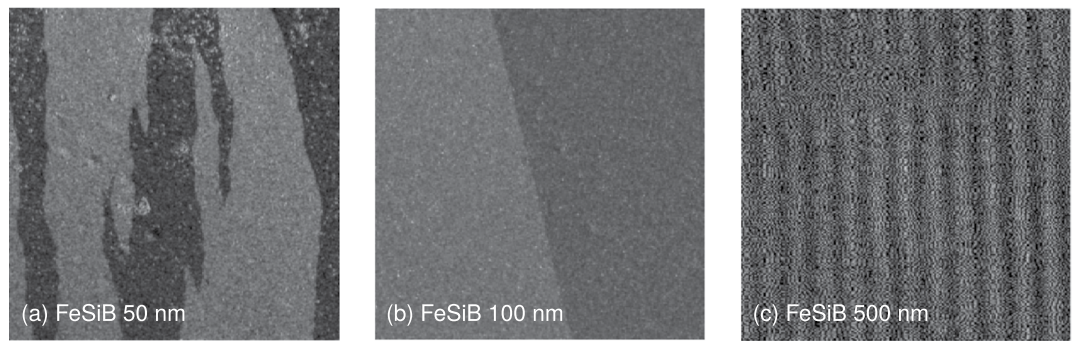

**Figure 6.** Evolution of the magnetic domain structure with thickness. Magnetic domain structure for amorphous FeSiB films with thicknesses of (**a**) 50, (**b**) 100 and (**c**) 500 nm. Films with thicknesses above ~150–200 nm have the same features observed for the 500-nm-thick film, which presents stripe magnetic domain structure, a configuration strictly related with isotropic in-plane magnetic properties and an out-of-plane anisotropy contribution[14,15,43,45,46,56–60]. Below this thickness, the films exhibit magnetic behavior of a classical in-plane uniaxial magnetic anisotropy system, without any out-of-plane anisotropy component[14,15,43,45,46,56–60]. Films with thickness between ~100 and ~150–200 nm present domain structure similar to that found for the 100-nm-thick film, characterized by large in-plane magnetic domains with antiparallel magnetization oriented along the easy axis. However, with decreasing thickness, we observe the emergence of domain walls with zigzag pattern, separating the in-plane magnetic domains, as evidenced for the 50-nm-thick film. In (a) and (b), the image size is $400 \times 400\ \mu m^2$, whereas it is $30 \times 30\ \mu m^2$ in (c). All images are taken at the remanence, after in-plane magnetic saturation. Specifically considering these images, the field is first applied along the vertical direction.

Why do the scaling behavior change with the film thickness? Noticeably, our results confirm that polycrystalline films have the old-plain DW dynamics governed by long-range interactions found for polycrystalline materials[29–31]. Hence, we interpret the change of the exponents and the evolution of the avalanche shape, found in polycrystalline films as the thickness is reduced from 100 to 50 nm, as a clear experimental evidence of a dimensional crossover in the DW dynamics, from three- to two-dimensional magnetic behavior.

Our results directly reveal that a dimensional crossover in the DW dynamics takes place within the thickness range between 100 and 50 nm for both, polycrystalline and amorphous films. But what makes this thickness range special? The thickness has fundamental role on the magnetic domain structure and DW formation, as well as it also affects the characteristics of the DW motion. Figure 6 shows representative domain images ilustrating the evolution of the magnetic domain structure with thickness. It is noteworthy that similar domain patterns have previously been reported for polycrystalline and amorphous samples with different compositions[15,43,56,58–60]. The thickness dependence of the magnetic properties and domain structure has been focus of many investigations in the last decades and, nowadays, despite the complexity of the issue, its main aspects are well understood[14,15]. For our set of samples, films with thicknesses above ~150–200 nm present stripe magnetic domain structure, a configuration strictly related with the isotropic in-plane magnetic properties and an out-of-plane anisotropy contribution[14,15,43,45,46,56–60]. Below this thickness, the films exhibit magnetic behavior of a classical in-plane uniaxial magnetic anisotropy system, without any out-of-plane anisotropy component, characterized by large in-plane magnetic domains with antiparallel magnetization oriented along the easy axis, separated by various types of domain walls strongly dependent on the film thickness[14,15,43,45,46,56–60]. However, although modifications of the magnetic domain structure are found within the thickness range between 100 and 50 nm, the magnetization in these films essentially lies in the plane, suggesting that the domain wall itself plays the major role here on the critical behavior of the magnetization process.

In contrast to bulk materials with relatively simple magnetic structure and nearly parallel DW, films show richer and often more complicated domains and DW patterns[12,13,30]. In soft ferromagnetic films with in-plane magnetic domains, despite the diversity of DW (Bloch walls, symmetric and asymmetric Néel walls, and the conspicuous cross-tie wall, this latter a complex pattern of Néel wall), the basic types of DW are simply Bloch and Néel walls. The type of domain wall will depend on the domain wall energy[61,62], which in turn for both wall types is dependent on the thickness, domain-wall thickness, effective magnetic anisotropy, saturation magnetization and exchange stiffness constant, or, in other words, is a result of the sum of the magnetostatic, exchange and anisotropy energy contributions[14,15,61,62]. Generally, the domain wall assumes the form of Bloch wall (in which an out-of-plane stray field exists in the domain wall due to the rotation of magnetic moments occurs in a perpendicular direction from the adjacent domains) when the film is thicker; and it will become Néel wall (in which the magnetic moments inside the wall strictly lie in the film plane, thus reducing the magnetostatic contribution to the wall energy) when the film thickness is below a critical value[14,15,61–63]. Actually, classical books[14,15] and reports addressing theoretically and experimentally the stability of DW in films[61–67] reveal that the well-known transition in which the domain wall passes from Bloch type to Néel type takes place in a critical thickness range between 100 and 50 nm. Hence, we understand that deviations from the critical behavior observed for the thickest films may be ascribed to the thickness, i.e., the smaller geometrical dimension of the system. Specifically, magnetic domains and domain walls are influenced by the film thickness due to the increasing importance of stray fields along the direction normal to the plane[12]. Between 100 and 50 nm, the thickness becomes of the same order of magnitude of the DW width and the stray fields constitute an appreciable source of magnetostatic energy, having straight impact on the inner structure of the DW[14,15,61,62] and, therefore, on the DW motion. Thereby, from the phenomenological point of view, the dimensional crossover may be seen as a consequence of this change in the type of DW, the Bloch-Néel transition. Due to the lost of one degree of freedom of the DW, with an essentially in-plane distribution of magnetic moments inside the wall, a two-dimensional description of the DW dynamics become reasonable for the films thinner than 50 nm.

So do we measure different exponents and avalanche shapes for polycrystalline and amorphous films? Yes, we do. It's natural to ask whether the established link found for bulk materials between microstructure of the materials and range of interactions[31] is still valid for films. Indeed, we confirm this relationship for polycrystalline films. Intuitively, one could expect that amorphous films irrespective of the thickness present a DW dynamics governed by short-range elastic interactions. This is particularly true for all films thicker than 100 nm, which share a common three-dimensional DW dynamics, despite of significant changes in the magnetic domain structure, as we can see in Fig. 6.

And, what happen with decreasing thickness? For thinner amorphous films though, the crucial agreement between experiment and theory reveals an unexpected critical behavior — films in two-dimensional regime naturally evolves towards a DW dynamics in which dipolar interactions are stronger than surface tension effects. Apparently, a crossover to an universality class describing two-dimensional DW dynamics with short-range interactions is only found when an external parameter, as temperature, is experimentally altered, thus tuning the scaling behavior according the dominant interaction in the system by modifying the DW structure[37].

The most striking finding here is that the change of exponents and avalanche shape for amorphous films reveals a crossover between two universality classes that is caused by both, change of system dimensionality and competition between the short-range DW surface tension and the long-range dipolar interaction. The interpretation that the dominant interaction changes from short-range to long-range interaction, simultaneously to the dimensional crossover, is consistent with the modification of the DW morphology[54] with decreasing thickness. Specifically, the contribution to the scaling behavior of strong long-range interactions of dipolar origin arises due to the appearance of the charged zigzag DW morphology[35–42,52–54] as the thickness is reduced from 100 nm, as we can confirm in Fig. 6. This report is the first to show the dimensional crossover in the DW dynamics and to disclose the outcomes of the interplay between system dimensionality and range of interactions governing the DW dynamics on Barkhausen avalanches.

The critical behavior in many systems can be explained by the range of interactions and system dimensionality. Theories and experiments are crucial to explain the signatures of the underlying avalanche dynamics, and they can help to uncover mysteries in a wide sort of systems. However, achieving a global perspective on the universality classes for crackling noise remains an open question. Inspired by numerous challenges in the field, we address here the crackling noise in ferromagnets. We believe that measuring an only power law is almost never definitive by itself. So we scrutinize the traditional statistics of Barkhausen avalanches in polycrystalline and amorphous ferromagnetic films having different thicknesses. Our results show how scaling exponents and average shape of the avalanches evolve with the structural character of the materials and film thickness, informing these features of the samples play fundamental role on the signatures of the underlying domain wall dynamics. Specifically, for films thicker than 100 nm, systems with three-dimensional magnetic behavior, scaling exponents vary according to the structural character of the sample, placing polycrystalline and amorphous materials in distinct universality classes associated with the kind and range of interactions governing the DW dynamics. Moreover, the exponents are dependent on the sample thickness, thus splitting thick and thin films into distinct classes, and inferring the need of a common two-dimensional description for films thinner than 50 nm, irrespective of the structural character. By comparing our experiments with theoretical predictions, we bring experimental evidence that supports the validity of several models for the DW dynamics. We also reveal that the films split into three well-defined universality classes of Barkhausen avalanches. Through the changes of the scaling exponents and avalanche shape, we observe the dimensional crossover in the DW dynamics and the outcomes of the interplay between system dimensionality and range of interactions governing the DW dynamics on Barkhausen avalanches. Thereby, we provide a clear picture to the crackling noise in magnetic systems with reduced dimensions. But of course the whole story is not over. After playing with universality classes of Barkhausen avalanches, we wonder how many systems throughout nature share similar interplay of fundamental features underlying crackling noise. So, let's play!

## Methods

**Ferromagnetic films.** We investigate Barkhausen avalanches in polycrystalline and amorphous ferromagnetic films with thicknesses from 20 to 1000 nm. The polycrystalline films have composition $Ni_{81}Fe_{19}$ (NiFe), whereas the amorphous alloys are $Fe_{75}Si_{15}B_{10}$ (FeSiB), $Co_{75}Si_{15}B_{10}$ (CoSiB), $Co_{77}Fe_{23}$ (CoFe), and $Fe_{73.5}Si_{22.5-x}Cu_1Nb_3B_x$ with $x = 4$ (B4), 6 (B6), and 9 (B9).

The films are deposited by magnetron sputtering onto glass substrates, with dimensions 10 mm × 4 mm, covered with a 2-nm-thick Ta buffer layer. The deposition process is carried out with the following parameters: base vacuum of $10^{-7}$ Torr, deposition pressure of 5.2 mTorr with a 99.99% pure Ar at 20 sccm constant flow, and DC source with current of 50 mA and 65 W set in the RF power supply for the deposition of the Ta and ferromagnetic layers, respectively. During the deposition, the substrate moves at constant speed through the plasma to improve the film uniformity, and a constant magnetic field of 1 kOe is applied along the main axis of the substrate in order to induce magnetic anisotropy.

**Structural and magnetic characterizations.** The structural characterization is obtained by x-ray diffraction. While low-angle x-ray diffraction is employed to determine the deposition rate and calibrate the film thickness, high-angle x-ray diffraction measurements are used to verify the structural character of each sample.

Quasi-static magnetization curves are obtained along and perpendicular to the main axis of the films, in order to verify the magnetic properties. Detailed information on the structural characterization and magnetic properties is found in refs[44–46,56].

To obtain further information on the magnetic behavior and magnetic domain morphology, images of the domain structure of the films are acquired by high resolution longitudinal Kerr effect experiments, on a $400 \times 400\,\mu m^2$ sample area, as well as by magnetic force microscopy, visualizing a $30 \times 30\,\mu m^2$ sample area. In particular, all images are taken at the remanence, after in-plane magnetic saturation.

**Barkhausen noise experiments.** We record Barkhausen noise time series using the traditional inductive technique in an open magnetic circuit, in which one detects time series of voltage pulses with a pickup coil wound around a ferromagnetic material submitted to a smooth, slow-varying external magnetic field, as we can see in Fig. 1(a). In our setup, sample and pickup coils are inserted in a long solenoid with compensation for the borders to ensure an homogeneous magnetic field on the sample. The sample is driven by a triangular magnetic field, applied along the main axis of the sample, with an amplitude high enough to saturate it magnetically. Here we perform experiments with driving field frequency in the range 0.05–0.4 Hz. Barkhausen noise is detected by a pickup coil (400 turns, 3.5 mm long and 4.5 mm wide) wound around the central part of the sample. A second pickup coil, with the same cross section and number of turns, is adapted in order to compensate the signal induced by the varying magnetic field. The Barkhausen signal is then amplified and filtered using a 100 kHz low-pass preamplifier filter, and finally digitized by an analog-to-digital converter board with sampling rate of $4 \times 10^6$ samples per second. Barkhausen noise measurements for all driving field frequencies are performed under similar experimental conditions. The time series are acquired just around the central part of the hysteresis loop, near the coercive field, where the DW motion is the main magnetization mechanism and the noise achieves the condition of stationarity[12,13,25]. In particular, for each ferromagnetic film, the following analyses are obtained from 200 time series.

**Statistical analysis of the Barkhausen avalanches.** Barkhausen noise is composed by a series of intermittent voltage pulses, i.e., avalanches, combined with background instrumental noise. At a pre-analysis stage, we employ a Wiener deconvolution, which optimally filters the background noise and removes distortions introduced by the response functions of the measurement apparatus in the original voltage pulses, thus obtaining reliable statistics despite the low intensity of the signal. Detailed information on the Wiener filtering is provided in ref.[44].

The following noise statistical analysis is performed using the procedure discussed in refs[21,31,44,68], in which a threshold is set to properly define the beginning and end of each Barkhausen avalanche. The inset in Fig. 1(b) shows an example of how the avalanches are extracted. The duration $T$ of the Barkhausen avalanche is estimated as the time interval between the two successive intersections of the signal with the threshold. The area underneath the avalanche signal, between the same points, is defined as the avalanche size $s$.

In contrast to magneto-optical techniques that restrict the analysis to the distribution of avalanche sizes, our experiments allow us to perform for films the wide statistical treatment usually employed for bulk materials. Here we identify the universality class of Barkhausen avalanches by measuring the distributions of Barkhausen avalanche sizes and avalanche durations, the average size as a function of the avalanche duration, power spectrum, and the average avalanche shape.

We observe that the measured $P(s)$, $P(T)$ and $\langle s \rangle$ vs. $T$ avalanche distributions typically follow a cutoff-limited power-law behavior and can be respectively fitted as

$$P(s) \propto s^{-\tau} e^{-(s/s_0)^{n_s}}, \qquad (1)$$

$$P(T) \propto T^{-\alpha} e^{-(T/T_0)^{n_T}}, \qquad (2)$$

$$\langle s \rangle \propto T^{1/\sigma \nu z} \left[ \frac{1}{1 + (T/T_0)^{n_{ave}(1/\sigma \nu z - 1)}} \right]^{1/n_{ave}}, \qquad (3)$$

where $\tau$, $\alpha$ and $1/\sigma \nu z$ are the scaling exponents, $s_0$ and $T_0$ indicate the position of the cutoff where the function deviates from the power-law behavior, and $n_s$, $n_T$, and $n_{ave}$ are the fitting parameters related to the shape of the cutoff function. In particular, we verify that the exponents are independent of the threshold level, at least for a reasonable range of values.

We observe that the measured $S(f)$ also follows a power-law behavior at the high frequency range of the spectrum, which can be described by[55]

$$S(f) \propto f^{-1/\sigma \nu z}. \qquad (4)$$

Although the power spectrum has not been considered for the fitting procedure, we confirm the theoretical prediction that the same scaling exponent can be employed to describe the power-law relationship between $\langle s \rangle$ and $T$, as well as the scaling regime of the power spectrum at high frequencies.

We go beyond scaling exponents and also focus on the average avalanche shape, a sharper tool for comparison between theory and experiments[1]. Here, we obtain both, the average temporal avalanche shape, considering the avalanches of a given duration $T$ and averaging the signal at each time step $t$, and the average avalanche shape for a given size (or magnetization), taking the avalanches of a size $s$ and averaging the signal at each size step $S$. The general scaling form for the average temporal avalanche shape, discussed in detail in ref.[4], is described by

$$\langle V(t|T) \rangle \propto T^{1/\sigma\nu z - 1} \left[ \frac{t}{T} \left( 1 - \frac{t}{T} \right) \right]^{1/\sigma\nu z - 1}, \qquad (5)$$

whereas the general scaling form for the avalanches of a given size is expressed as

$$\langle V(S|s) \rangle \propto s^{1-\sigma\nu z} \left[ \frac{S}{s} \left( 1 - \frac{S}{s} \right) \right]^{1-\sigma\nu z}. \qquad (6)$$

Remarkably, the exponent $1/\sigma\nu z$ characterizes the scaling regime and leads to an evolution of the average avalanche shape with the universality class.

### Acknowledgements
F.B. would like to thank S. Zapperi for enlightening discussions. F.B., M.A.C., N.R.M., R.D.D.P., C.C. and R.L.S. acknowledge financial support from the Brazilian agencies CNPq and CAPES. G.D. acknowledges the support of PSL Grant No. ANR-10-IDEX-0001-02-PSL.

### Author Contributions
F.B. was responsible for the experiments and analysis of the Barkhausen avalanches. All authors interpreted the results. F.B. wrote the original text of the manuscript. All authors contributed to improve the text.

### Additional Information
**Competing Interests:** The authors declare no competing interests.

**Publisher's note:** Springer Nature remains neutral with regard to jurisdictional claims in published maps and institutional affiliations.